\documentclass[doublespacing,dvips,dvipdfm]{elsart}
\journal{Physics Letters B}

\usepackage{graphicx}
\usepackage{color}
\begin{document}

\begin{frontmatter}
\title{Low-lying Proton Intruder State in $^{13}$B}
\author[cns]{S.~Ota}
\author[cns]{S.~Shimoura}
\author[tokyo]{H.~Iwasaki}
\author[riken]{M.~Kurokawa}
\author[cns]{S.~Michimasa}
\author[riken]{N.~Aoi}
\author[riken]{H.~Baba}
\author[rikkyo]{K.~Demichi}
\author[atomki]{Z.~Elekes}
\author[riken]{T.~Fukuchi}
\author[riken]{T.~Gomi}
\author[riken]{S.~Kanno}
\author[cns]{S.~Kubono}
\author[rikkyo]{K.~Kurita}
\author[rikkyo]{H.~Hasegawa}
\author[cns]{E.~Ideguchi}
\author[tohoku]{N.~Iwasa}
\author[rikkyo]{Y.U.~Matsuyama}
%\author[msu]{K.~Miller}
\author[msu]{K.L.~Yurkewicz}
\author[riken]{T.~Minemura}
\author[riken]{T.~Motobayashi}
\author[kyoto]{T.~Murakami}
\author[anl]{M.~Notani}
\author[osaka]{A.~Odahara}
\author[tokyo]{A.~Saito}
\author[riken]{H.~Sakurai}
\author[riken]{E.~Takeshita}
\author[riken]{S.~Takeuchi}
\author[cns]{M.~Tamaki}
\author[kyushu]{T.~Teranishi}
\author[riken]{Y.~Yanagisawa}
\author[riken]{K.~Yamada}
\author[riken]{M.~Ishihara}

\address[cns]{Center for Nuclear Study, University of Tokyo, Saitama
 351-0198, Japan}
\address[tokyo]{Department of Physics, University of Tokyo, Tokyo
 113-0033, Japan}
\address[riken]{RIKEN Nishina Center for Accelerator-Based Science,
 Saitama 351-0198, Japan}
\address[rikkyo]{Department of Physics, Rikkyo University, Tokyo
 171-8501, Japan}
\address[atomki]{Hungarian Acad. Sci., Inst. Nucl. Res., H-4001 Debrecen,
 Hungary}
\address[tohoku]{Department of Physics, Tohoku University, Miyagi
 980-8578, Japan}
\address[msu]{NSCL, Michigan State University, East Lansing, MI 48824,
 USA}
\address[kyoto]{Department of Physics, Kyoto University,  Kyoto
 606-8502, Japan}
\address[anl]{Argonne National Laboratory, Argonne, IL 60439, USA}
\address[osaka]{Department of Physics, Osaka University 560-0043, Japan}
\address[kyushu]{Department of Physics, Kyushu University, 812-8581}
\begin{abstract}

% Text of abstract
The neturon rich nucleus $^{13}$B was studied via the proton transfer reaction
 $^{4}$He($^{12}$Be,$^{13}$B$\gamma$) at 
 50$A$~MeV.  The known 4.83-MeV excited state was strongly populated and
 its spin and parity were assigned to 1/2$^{+}$ by comparing the
 angular differential cross section data with DWBA calculations.  
 This low-lying 1/2$^{+}$ state is interpreted as a proton intruder
 state and indicates a  deformation of the nucleus.
% A dynamical change of the proton shell structure is discussed.
\end{abstract}

\begin{keyword}
 Intruder state \sep 
 Proton single particle state \sep Proton transfer reaction \sep
 $^{4}$He($^{12}$Be,$^{13}$B$\gamma$) \sep 
% 21.10.Pc 	Single-particle levels and strength functions
% 25.55.Hp 	Transfer reactions
% 27.20.+n 	6(less-than-or-equal-to)A(less-than-or-equal-to)19
% 29.30.Kv 	X- and gamma-ray spectroscopy
\PACS 21.10.Pc \sep 25.55.Hp \sep 27.20.+n \sep 29.30.Kv
\end{keyword}

\end{frontmatter}

\maketitle
%\section{Introduction}
% keyword
%
%
The existence of intruder states in light neutron-rich unstable nuclei is
often considered to be evidence for one or more $\hbar\omega$
configurations in the low-lying states around the $psd$ shell.
The ground state of $^{11}$Be is $1/2^{+}$ which is lower in energy by
0.3~MeV than the $1/2^{-}$ state \cite{Wilkinson1959}.
In $^{12}$Be, there is a $1^{-}$ intruder
state at 2.7-MeV excitation energy \cite{Iwasaki2000}.
The energies of these low-lying, non-normal parity states indicate
$1\hbar\omega$ configurations.
Furthermore, the presence of low-lying 2$_{1}^{+}$ \cite{Iwasaki2000a}
and 0$_{2}^{+}$ \cite{Shimoura2003,Shimoura2007} states in $^{12}$Be
suggests a $2\hbar\omega$ configuration.

Three theoretical interpretations have been proposed for 
these one or more $\hbar\omega$ configurations in the low-lying states of
neutron-rich unstable nuclei:
(1) the monopole interaction of the tensor force \cite{Suzuki2003},
(2) the loosely bound nature of some orbitals \cite{Hamamoto2004}, and
(3) nuclear deformation \cite{Shimoura2007,BohrMottelson,Hamamoto2007}.
In Ref. \cite{Suzuki2003},the effective interaction was determined so
that the model reproduces the energy levels in light, neutron-rich
nuclei including intruder states,
and the importance of monopole interaction due to the tensor
force was pointed out.
Reference \cite{Hamamoto2004} discusses the fact that in the $psd$
shell, the 
$2s_{1/2}$ orbital gains its energy relative to the other orbitals due
to its loosely bound nature.
The non-zero $\hbar\omega$ configurations
can also be intuitively explained by the deformed mean field picture.
As seen in the Nilsson diagram, the gap becomes smaller with increasing
deformation \cite{Shimoura2007,BohrMottelson,Hamamoto2007}.
Since the combination of these effects, which are provided by their
corresponding theoretical models, changes the neutron shell 
structure in neutron-rich nuclei, such as $^{11,12}$Be,
the relative importance of the three theoretical approaches has not been
clarified.

For the proton shell in light neutron-rich
nuclei, the effects due to the tensor force and the loosely bound
nature of the orbitals are expected to be small since the $\nu p_{1/2,3/2}$
orbitals are fully filled and the proton(s) are deeply bound.
However, deformation is still presumed to affect the proton shell
structure. 
Proton intruder states are, therefore, signatures for the importance
of deformation.
In the present study, we focus on the proton shell structure in the
$N=8$ nucleus $^{13}$B by investigating the proton single-particle
states. 

In order to investigate the proton single-particle states in $^{13}$B,
we used the $(\alpha, t)$ reaction on $^{12}$Be in inverse kinematics. 
This process at an intermediate energy has a relatively
large cross section since the proton to be transferred is
deeply bound in the $\alpha$ particle and, thus, has high-momentum
components, which reduce the effect 
of the momentum mismatch of the ($\alpha$,$t$) reaction \cite{Michimasa2006}. 
%We observed a large cross section of 4.83-MeV excitation
%and assigned the spin and parity of $1/2^{+}$ to the 4.83-MeV excited
%state, which is a proton intruder state.

The experiment was performed at the RIKEN Accelerator Research Facility.
A $^{12}$Be beam was produced by
the fragmentation reaction of a 100$A$~MeV $^{18}$O beam on a
$^{9}$Be target with a thickness of 1.85-g/cm$^{2}$.
The $^{12}$Be beam was separated by the 
RIKEN Projectile-fragment Separator (RIPS) \cite{Kubo1992}.
The incident particles were identified on an event-by-event basis using the
measured time-of-flight and the energy deposited.
The time-of-flight, over a path length of 5.4~m, and the energy
deposited were measured using two plastic scintillators located at the
last two foci of the RIPS.
The intensity and the purity of the $^{12}$Be beam respectively were
typically 2 $\times$ 10$^{5}$ counts per second and 90\%.

\begin{figure}[H]
 \begin{center}
  \includegraphics[width=10cm]{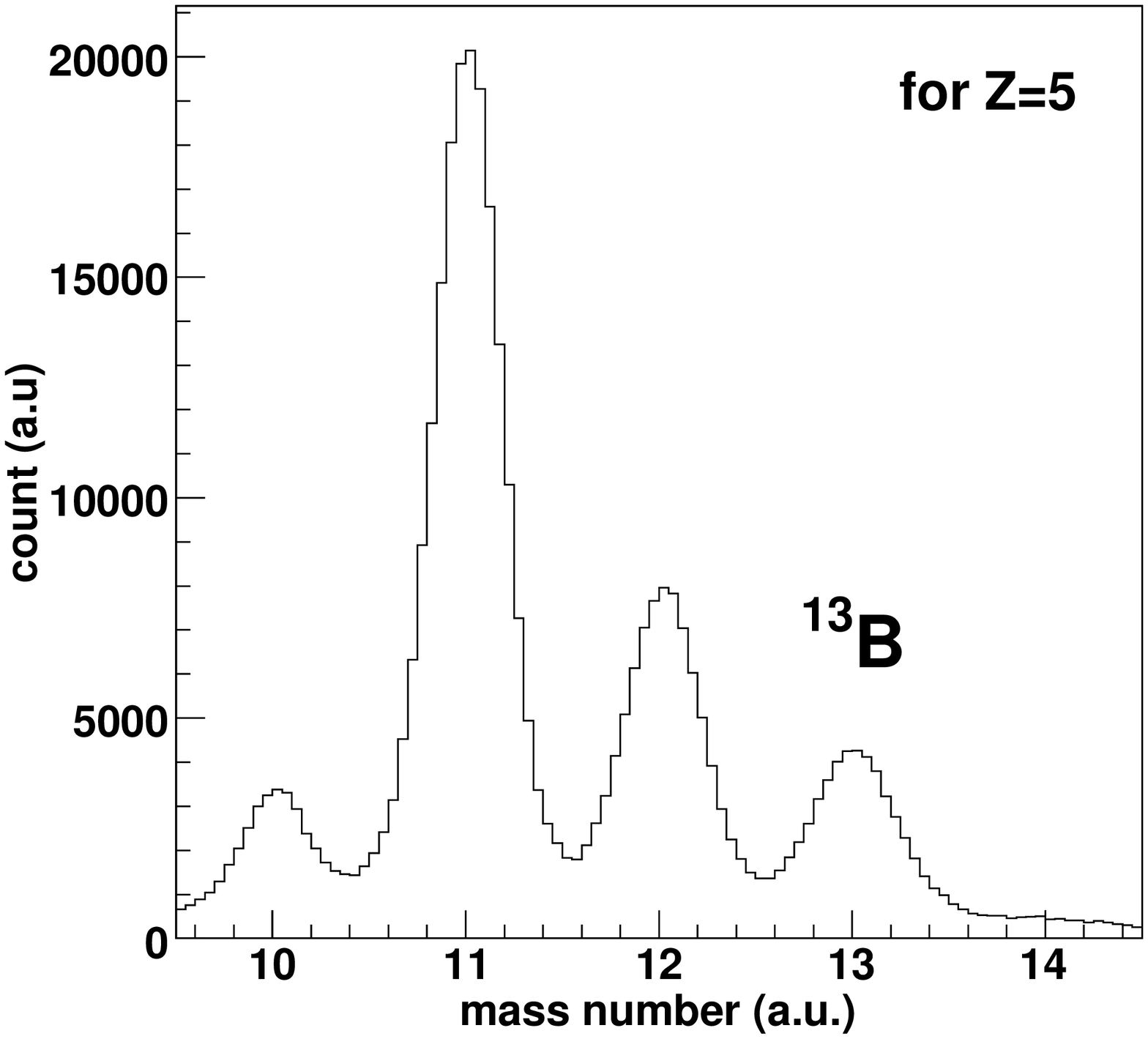}
 \end{center}
 \caption{Mass distribution of the outgoing particles for $Z=5$
 extracted from the correlation between TOF and $\Delta E$ measured with
 the plastic scintillator hodoscope.
 \label{fig:mass-distribution}}
\end{figure}

The 50$A$ MeV $^{12}$Be beam bombarded a secondary target of
liquid helium\cite{Ryuto2005} located at the final focus of the RIPS.
A liquid helium target was chosen because of its statistical advantage
in terms of the experimental yields.
The helium was condensed by a cryogenic refrigerator and kept below 4~K
during the experiment.
A target thickness of 143$\pm$5~mg/cm$^2$ was estimated from the
velocity difference between outgoing particles measured with and
without the liquid helium. 
The positions and directions of the incident particles
on the secondary target were deduced from the position information of
two parallel plate avalanche counters \cite{Kumagai2001} located
30-cm apart from each other around the final focus.
The outgoing particles were detected by a plastic scintillator
hodoscope 3.5-m downstream from the secondary target with
a 1~$\times$~1 m$^2$ active area and an angular 
acceptance of up to 8 degrees in the laboratory frame.
%corresponding to 39 degrees in the center-of-mass frame.  
The plastic scintillator hodoscope consisted of 5-mm thick $\Delta E$
and 60-mm thick $E$ layers.
The $\Delta E$ layer was divided into 13 plastic scintillator bars
vertically and the $E$ layer into 16 plastic scintillator
bars horizontally. 
The outgoing particles were identified on an event-by-event
basis using the measured time-of-flight and the energy deposited in
the $\Delta E$ and $E$ layers.
The mass distribution for the $Z=5$ particles is shown in
Fig.~\ref{fig:mass-distribution}; the
mass resolution $(\delta A)$ was determined to be $\sim$0.25.
The time-of-flight between the secondary target and the hodoscope was
deduced from the time information of the plastic scintillators located
upstream from the secondary target and the plastic scintillator
hodoscope.
Position information for the outgoing particles was 
deduced from the time difference between the output signals from the two
photomultiplier 
tubes attached to both ends of each scintillator bar, and was used
to determine the scattering angle.
The angular resolution of the scattering angle in the laboratory frame
was 0.5 degrees in one standard deviation.

The de-excitation $\gamma$ rays were detected by an array of germanium
detectors: Gamma-Ray detector Array with
Position and Energy sensitivities (GRAPE)\cite{Shimoura2004}.  This
consisted of 6 germanium detectors located at 140$^\circ$ with respect
to the beam axis.  Each detector contains two cylindrical crystals 
6~cm in diameter and 2~cm thick, with a common  anode between
them.
The each cathode attached to each crystal is segmented into a 
$3 \times 3$ matrix.
The GRAPE provides position information of the
$\gamma$-ray interaction point, which is extracted from a pulse shape
analysis of the signal from the cathode
\cite{Shimoura2004,Kurokawa2003}. 
The intrinsic energy resolution and the full energy peak efficiency were
typically 2.7~keV (FWHM) and 0.4\%, respectively, for
1332-keV $\gamma$ rays from a $^{60}$Co standard source.  
The energy resolution 
after correcting for the Doppler shift was deduced to be 1.3\% (FWHM)
for 2.1-MeV $\gamma$ rays, corresponding to the decay of the first
2$^{+}$ state of $^{12}$Be moving with 30\% the light velocity.  
The excited states of $^{13}$B populated in the reaction were identified
by measuring the energy of the de-excitation $\gamma$ rays.

\begin{figure}[H]
 \begin{center}
  \includegraphics[width=15cm]{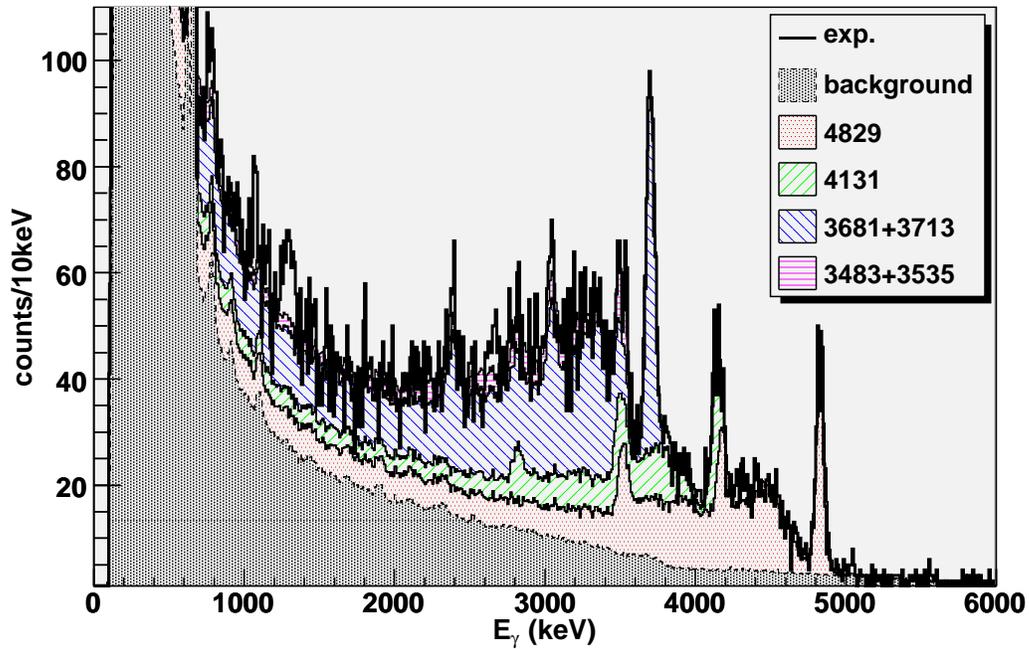}
 \end{center}
 \caption{
 Doppler corrected $\gamma$-ray spectrum in coincidence with  $^{13}$B
 particles.
 The spectrum is decomposed with the sum of response functions of the
 GRAPE for the de-excitation and background $\gamma$ rays.
 The resulting responce functions are shown with hatched areas.
 The response functions for two doublets around 3.5 and 3.7~MeV are
 summed.
 The response function for the background includes the natrural
 background and the decay of the $^{12}$Be isomer (see the text).
 \label{fig:energyspectrum}}
\end{figure}

Figure \ref{fig:energyspectrum} shows the energy spectrum of $\gamma$ 
rays measured in coincidence with $^{13}$B after correcting for the
Doppler shift.
There are three peaks corresponding to the transitions from (3.68-,
3.73-), 4.13- and 4.83-MeV excited states to the ground state.
The peaks seen in the low energy region is considered to originate
from the reaction of the beam and the window of the target cell.
Other significant transitions including those between the
excited states were not observed in the present measurement.
In the figure, hatched areas show the response functions
of the GRAPE for the de-excitation $\gamma$ rays obtained by means of a
Monte Carlo simulation using the  GEANT4 
code \cite{GEANT2003,GEANT4WWW} and for the background $\gamma$ rays.
The background was assumed to consist of two components, the natural
background $\gamma$ rays, estimated by
putting the gate in the non-prompt region of the time spectrum of
the GRAPE,
and the $\gamma$ rays from the isomer state of $^{12}$Be, which
were simulated by assuming a life time of 331~ns \cite{Shimoura2007}.
The ratio of the isomer state to the ground state in the secondary beam
was less than 2 percent at the secondary target.
In the simulations, all the resolutions of the detectors
associated with the correction for the Doppler shift were taken into
account.
The intensity of each decay to the ground state was deduced by fitting
the experimental spectrum with the sum of the response functions and the
background. 
Assuming no cascade decay, the derived relative intensities of the observed
$\gamma$ rays from the known excited states are shown in
Table~\ref{tab:cross-sections}, together with the previously reported
data including two neutron transfer \cite{Ajzenberg-Selove1978}, 
neutron knockout \cite{Guimaraes2000}, 
$\beta$-decay followed by neutron decay \cite{Aoi2002}, 
and multi-nucleon transfer \cite{Kalpakchieva2000}.
In the present reaction, the excited state at 4.83 MeV is strongly
populated relative to the other excited states,
while it is less excited in the other reactions except for the
multi-nucleon transfer reaction which is expected to populate proton
excited states. 
Considering the selectivity of the proton transfer
reaction, it is conceivable that the 4.83-MeV excited state is of proton
single-particle nature.

\begin{figure}[H]
 \begin{center}
  \includegraphics[width=7.5cm]{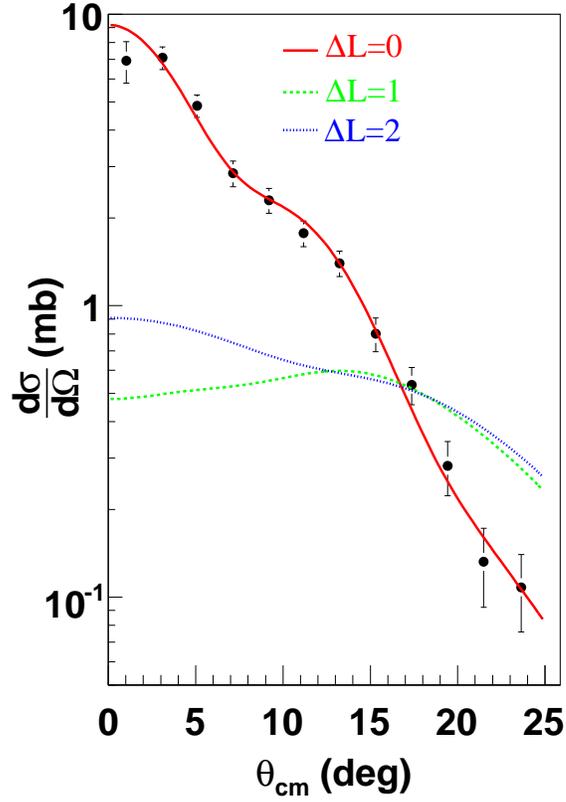}
 \end{center}
 \caption{Angular distribution of the 4.83-MeV excited $^{13}$B. 
 Experimental data is shown by the filled circles with statistical
 errors. The solid, dotted and dashed curves show the DWBA predictions
 with assumptions of $\Delta L$~=~0, 1, and 2, respectively. 
 See the text for more detail including the optical potentials.
}\label{fig:angularDistribution}
\end{figure}

In order to determine the angular differential cross section of the
$^{4}$He($^{12}$Be,$^{13}$B$^*$) reaction, 
individual $\gamma$-ray spectra with 0.5-degree-scattering-angle cuts
were fitted with the same response functions and background described
above, and they were analyzed to deduce the populations.
Figure~\ref{fig:angularDistribution} shows the experimental angular
distribution for the 4.83-MeV excitation with filled circles.
The transferred angular momentum ($\Delta L$) in the reaction is
determined by comparing the obtained angular distributions with
predictions of the distorted-wave Born approximations (DWBA), calculated with
the DWUCK5 code \cite{Kunz}.
The optical potentials for the entrance and exit channels were obtained by
adopting a single folding model used in Ref.~\cite{Satchler1997}.
The density distribution of $^{12}$Be for the folding is calculated by
using the mean field calculation code, TIMORA\cite{Horowitz1991,Horowitz1981}.
The density distribution of $^{13}$B is assumed to be the same as that
of $^{12}$Be with $R(^{13}{\rm B})/R(^{12}{\rm Be})=(13/12)^{1/3}$.
For the entrance channel, the depth of the imaginary potential is 
adjusted so as to represent the inelastic scattering data of $^{12}$Be,
excited to the $2^{+}_{1}$ state.
For the exit channel, the depths of the real and imaginary potentials
are varied in order to estimate the statistical error for the
spectroscopic factor. 
The DWBA predictions with $\Delta L=0, 1, 2$ are shown as curves in
Fig.~\ref{fig:angularDistribution}.
The absolute magnitudes of the predictions are normalized so as to
minimize the $\chi^{2}$ values.
The reduced $\chi^{2}$ value is 0.74 for the $L=0$ calculation.
The forward angle peak in the experimental angular distribution is well
described by the $\Delta L$~=~0 DWBA calculation. 
Therefore, we assigned a spin and parity of 1/2$^{+}$ to the 4.83-MeV
excited state. Its low energy indicates that it is a
proton intruder state from the $sd$ shell.
The spectroscopic strength $C^{2}S$ was deduced to be
$0.20\pm0.02$, where $S$ is a spectroscopic factor
and $C^{2}$ is an isospin Clebsch-Gordan coefficient. 
The systematic errors of the $C^{2}S$ due to the choice of the
optical potentials in the DWBA calculation and to the geometrical
uncertainty of the GRAPE in the simulation were evaluated to be 50\%
and 10\%, respectively.
The former one is obtained from the difference between the folding model
and a separate calculation using a global optical potential for
$^{12}$C in Ref.~\cite{Ingemarsson2000,Ingemarsson2001}.
In the absolute magnitude of the differential cross section,
there is no ambiguity originated from cascade decays to or from the
4.83-MeV excited state.
No excited state was found above 4.83~MeV up to the neutron
threshold of 4.87~MeV in $^{13}$B, therefore, the cascade decays to the
4.83-MeV excited state are unlikely to occur.
On the other hand, the cascade decay from the 4.83-MeV excited state is
also small  as explained below.
The excited state (1/2$^{+}$) is expected to decay to the ground state
(3/2$^{-}$) via an $E1$ transition, whose decay rate is proportional to
the cube of the transition energy, {\it i.e.},
$E_{\gamma}^{3}$ \cite{BohrMottelson}.
The maximum possible transition energy to a known excited state is
1.4~MeV corresponding to the decay to the 3.48-MeV excited state.
Even if the decay to this state is by $E1$ transition, 
its decay rate is about 40 times smaller than to the ground state.
Decay rates other than $E1$ are also expected to be much smaller.
In fact, no cascade decay of the 4.83-MeV excited state was observed in
the present and the previous experiments.

There are few theoretical predictions for $^{13}$B.
In Ref.~\cite{Guimaraes2000}, the result of a shell model calculation for
$^{13}$B has been shown, however, the calculation is performed from the
viewpoint of the neutron structure.
Focusing on the proton shell structure, we performed 
a shell model calculation for $^{13}$B($^{12}$Be)  with the
calculation code, OXBASH \cite{BrownWWW},
wherein the interaction including the effect of the
tensor force and reproducing intruder states in
the neutron-rich nuclei such as $^{11,12}$Be \cite{Suzuki2003}, is used.
The model space consisting of the $psd$ shell with maximum 3(2)$\hbar
\omega$ excitation was considered for $^{13}$B($^{12}$Be).
The spectroscopic factor was calculated as the overlap between each
excited state of $^{13}$B and the ground state of $^{12}$Be.
The shell model calculation predicts four 1/2$^{+}$ states below 10~MeV
excitation energy. 
Three lower states are predicted at 3.7, 6.5 and 9.1~MeV with
spectroscopic strengths of 0.01, 0.03 and 0.003,
respectively.
These strengths are too small to explain the experimental result. 
The fourth 1/2$^{+}$ state with a spectroscopic strength of 0.30 is
close to the experimental spectroscopic strength; however its
excitation energy is 9.5~MeV.
These results of the shell model calculation may indicate that 
the theoretical approach based on the effect of the tensor force does
not explain the experimental result,
or the model space consisting of the $psd$ shell and the excitation of
3$\hbar \omega$  are not large enough.

A simple expression for the observed $1/2^{+}$ state with proton single
particle nature is $(\pi sd)^{1} \otimes ^{12}$Be.
Concerning the possible large deformation of $^{12}$Be, 
we examine the qualitative consequence of a deformed core.
The $1/2^{+}$ state arises with a
configuration of $(\pi[220\frac{1}{2}])^{1}$ to the $^{12}$Be(g.s.)
core.
As seen in the Nilsson diagram, at large deformation, the
$\pi[220\frac{1}{2}]$ orbital gains energy.
The energy gain of the total system compared to the spherical $s_{1/2}$ is
estimated to be around 7~MeV from the diagram and the
semi-empirical mass formula with a deformation parameter of 0.5.
The discrepancy between the shell model prediction mentioned above and
the present result for the $1/2^{+}$ state is thus ascribed to a core
deformation effect.
In the present reaction, the observed $1/2^{+}$ state is
considered to be populated by the transfer of a proton to the
$\pi[220\frac{1}{2}]$ orbital. 
Hence, the low excitation energy of the 1/2$^{+}$ state suggests
a deformed mean field.
Since the proton is transfered to the $[220\frac{1}{2}]$ orbital by
an $s$-wave ($\Delta L = 0$),
the spectroscopic factor of the spherical $2s_{1/2}$ component in the
$[220\frac{1}{2}]$ orbital, {\it e.g.} $S=0.273$ at
$\delta=0.4$ \cite{BohrMottelson}, may also reproduce the experimental
result.

Based on the discussion above, the $1/2^{+}$ intruder state at 4.83~MeV
is a signature of a deformed field and indicates the importance of
a deformation in this nuclei.  On the other hand, the ground state of
$^{13}$B is supposed to be spherical with a normal $p$-shell
configuration \cite{Nagatomo2004a}.
The excitation of only one proton thus changes the structure of
the nucleus.

Recently an AMD calculation related to the experimentally assigned
1/2$^{+}$ state of $^{13}$B was carried out by Kanada-En'yo et al
\cite{Kanada2007}. 
The calculation predicts a largely deformed 1/2$_{1}^{+}$ state with
proton $1\hbar \omega$ excitation and neutron $2\hbar \omega$
excitation.
In the excited state, the last proton occupies a molecular $\sigma$ orbital
which has a density distribution similar to the $[220\frac{1}{2}]$
Nilsson orbital.
The 1/2$_{1}^{+}$ state predicted by AMD may correspond to the experimental
one although the calculated excitation energy is 8~MeV.

In summary, we have studied the proton transfer reaction on $^{12}$Be in
inverse kinematics.  
A spin and parity of $1/2^{+}$ for the 4.83-MeV excited state in
$^{13}$B were assigned for the first time.  
A spectroscopic strength of  $0.20\pm0.02$  was also obtained with
60\% systematic error.
This state is interpreted to be  a proton intruder state from the
$sd$-shell 
because of its non-normal parity and small excitation energy.
The nuclear deformation provides a simple mechanism for the existence of
such a low-lying proton intruder state.
The present study shows the importance of deformation for
proton shell structure in neutron-rich nuclei and 
the change of the proton shell structure deu to the excitation of one
proton.

We would like to thank the RIKEN Ring Cyclotron staff members for their
operation during the experiment. The present work was partially
supported by the Grant-in-Aid for Scientific Research (No.15204018) by
the Ministry of Education, Culture, Sports, Science and Technology, of
Japan.

\begin{table}[H]
  \caption{List of the relative populations of the excited states. 
 The intensities are normalized by the most intense population. The
 errors are statistical only.
 In the last column, the spins and parities which have been assigned in
 the previous and present studies are listed.
 a:Present reaction of $^{4}$He($^{12}$Be,$^{13}$B$\gamma$).
 b:$^{11}$B (t,p)$^{13}$B.
 c:$^9$Be ($^{14}$B,$^{13}$B)X.
 d:$^{14}$Be $\beta$ delayed n.
 e:$^{16}$O($^{14}$C,$^{17}$F)$^{13}$B.
  For the reaction b the relative intensity is at $\theta_{\rm lab}=10^{\circ}$ and 
 for the reaction e at $\theta_{\rm c.m.}=5.4^{\circ}$.
 }\label{tab:cross-sections}
 \begin{center}
  \begin{tabular}{ccccccc}
   \hline
   \hline
   & \multicolumn{5}{c}{Relative intensities} & \\
   E$_{x}$ (MeV)     & a & b & c & d & e & $J^\pi$\\
   \hline
   3.48      & 0.19 $\pm$ 0.05 & 0.06      & 0.60 $\pm$ 0.14 &    &    &  \\
   3.53      & 0.20 $\pm$ 0.05 & 0.19      &                 & 1  &    &  \\
   3.68      & 0.74 $\pm$ 0.07 & 0.38      & 1               &    &    &  \\
   3.71      & 0.68 $\pm$ 0.07 & 0.25      &                 &    &    &  \\
   4.13      & 0.49 $\pm$ 0.04 & 1         & 0.04 $\pm$ 0.04 &    &    &  \\
   4.83      & 1               & 0.03      &                 &    & 1  & 1/2$^{+}$ \\
   \hline
   \hline
  \end{tabular}
 \end{center}
\end{table}
\end{document}